\documentclass[sigconf,natbib=true,anonymous=false]{acmart}
\usepackage{latexsym}
\usepackage{subcaption}
\usepackage{adjustbox}
\usepackage{multirow}
\usepackage{graphics}
\usepackage{microtype}
\usepackage{amsmath}
\usepackage{bm}
\usepackage{graphicx}
\usepackage{xcolor}
\usepackage{array}
\usepackage{amsmath}
\usepackage{balance}

\AtBeginDocument{%
  \providecommand\BibTeX{{%
    \normalfont B\kern-0.5em{\scshape i\kern-0.25em b}\kern-0.8em\TeX}}}

\setcopyright{acmcopyright}
\copyrightyear{2022}
\acmYear{2022}
\setcopyright{rightsretained}
\acmConference[SIGIR '22]{Proceedings of the 45th International ACM SIGIR Conference on Research and Development in Information Retrieval}{July 11--15, 2022}{Madrid, Spain}
\acmBooktitle{Proceedings of the 45th International ACM SIGIR Conference on Research and Development in Information Retrieval (SIGIR '22), July 11--15, 2022, Madrid, Spain}\acmDOI{10.1145/3477495.3531860}
\acmISBN{978-1-4503-8732-3/22/07}

\usepackage{etoolbox}
\makeatletter
\patchcmd{\maketitle}{\@copyrightpermission}{
   \begin{minipage}{0.3\columnwidth}
     \href{http://creativecommons.org/licenses/by/4.0/}{\includegraphics[width=0.90\textwidth]{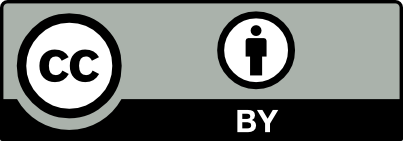}}
   \end{minipage}\hfill
   \begin{minipage}{0.7\columnwidth}
     \href{http://creativecommons.org/licenses/by/4.0/}{This work is licensed under a Creative Commons Attribution International 4.0 License.}
   \end{minipage}

   \vspace{5pt}
}{}{}

\makeatother

\begin{document}
\fancyhead{}
\title{Long Document Re-ranking with Modular Re-ranker}


\author{Luyu Gao}
\affiliation{ \vspace{0.1cm}
    Language Technologies Institute\\
    Carnegie Mellon University\\
    \city{Pittsburgh}
  \state{PA}
  \country{USA}
  \postcode{15213}\\
}
\email{luyug@cs.cmu.edu}

\author{Jamie Callan}
\affiliation{ \vspace{0.1cm}
    Language Technologies Institute\\
    Carnegie Mellon University\\
    \city{Pittsburgh}
  \state{PA}
  \country{USA}
  \postcode{15213}\\
}
\email{callan@cs.cmu.edu}

\renewcommand{\shortauthors}{Gao and Callan}

\begin{abstract}
  Long document re-ranking has been a challenging problem for neural re-rankers based on deep language models like BERT. Early work breaks the documents into short passage-like chunks. These chunks are independently mapped to scalar scores or latent vectors, which are then pooled into a final relevance score. These encode-and-pool methods however inevitably introduce an information bottleneck: the low dimension representations.
 In this paper, we propose instead to model full query-to-document interaction, leveraging the attention operation and modular Transformer re-ranker framework. First, document chunks are encoded independently with an encoder module. An interaction module then encodes the query and performs joint attention from the query to all document chunk representations. We demonstrate that the model can use this new degree of freedom to aggregate important information from the entire document. Our experiments show that this design produces effective re-ranking on two classical IR collections Robust04 and ClueWeb09, and a large-scale supervised collection MS-MARCO document ranking.\footnote{Our code is available at \url{https://github.com/luyug/mores_plus}.}
\end{abstract}

\maketitle

\section{Introduction}
Deep pre-trained Transformer~\cite{Vaswani2017AttentionIA} language models~(LM) like BERT~\cite{Devlin2019BERTPO} have been widely adopted in search and re-ranking and achieve state-of-the-art performance~\cite{yang2019xlnet,nogueira2019passage,dai2019deeper}. These models are composed of Transformer layers, which use self-attention operations~\cite{Vaswani2017AttentionIA} to contextualize and interact query and document~\cite{nogueira2019passage,dai2019deeper}. 

Despite their success, Transformer LMs have pre-set input length limits, typically 512, due to complexity considerations~\cite{Devlin2019BERTPO,liu2019roberta,yang2019xlnet}. As a result, re-ranking long documents beyond this length has been a challenging problem. Prior research designed techniques to accommodate the neural LM. For example, some attempts to strategically limit the input length, which however prohibits full interaction among all query and document tokens. Others have turned to more efficient Transformers that use lower complexity but also lower capacity sparse self-attention. In this paper, we present an alternative method, re-ranking documents with a modular re-ranker and a new form of attention operation, query to document cross attention.

Modular re-rankers such as MORES~\cite{gao-etal-2020-modularized} provide an alternative framework to common re-rankers. With MORES, query and document are independently encoded. A Transformer Interaction Module then makes relevance estimates using light-weight query-to-document attention. In this paper, we introduce a modification to this framework, MORES+, for long document re-ranking. We break a candidate document into passage-like chunks and encode them separately using identical Transformer encoders. A joint \emph{query-to-all-chunk} attention then models interaction between query and the \emph{full} document. The model can flexibly control the attention weights to pick up important information. This design roughly resembles the human reading comprehension process: after browsing the article and the question, the reader will refer back to the relevant pieces in the original article to come up with answers.

We present experiments with classical few-shot document IR datasets, ClueWeb09 and Robust04, as well as a recent large-scale dataset, MS-MARCO document ranking. We found that the MORES+ re-ranker produces better ranking quality on detailed natural language queries where interaction between query and document is more sophisticated.

The rest of the paper is organized as follows. Section 2 provides a background on document re-ranking methods with Transformers. Section 3 describes the MORES modular re-ranker and how we generalize it to MORES+ for document re-ranking. Section 4 presents experimental results with MORES+.

\section{Background}

The revolution in pre-trained Transformer language models~\cite{Peters:2018,Devlin2019BERTPO} had a large impact on IR~\cite{dai2019deeper}. Encoder-only LMs like BERT~\cite{Devlin2019BERTPO} are the current state-of-the-art for text re-ranking. They can a) generate contextualized representations~\cite{Peters:2018} to help enhance the language understanding capability of the re-rankers and b) perform deep interactions between query and document~\cite{dai2019deeper}. Other forms of pre-trained LMs also exist. Encoder-decoder LMs are pre-trained on sequence-to-sequence tasks and have previously been used in IR for tasks like document expansion\cite{Raffel2019ExploringTL,Lewis2020BARTDS}.

Despite their advantages, Transformer LM re-rankers cannot be naively used for long document ranking since most pre-trained LMs have a maximum input length constraint of 512. Prior research circumvented this by splitting documents into passage-like short text chunks. The passages are first separately judged and passage-level results are further pooled to generate the final prediction. Score-pooling systems simply take max/sum/first of the chunk scores~\cite{dai2019deeper}. These systems are typically named after their underlying LM and pooling method. For example, BERT-maxp~\cite{dai2019deeper} refers to a BERT-based system with max pooling. A generalization is representation aggregation systems like PARADE~\cite{Li2020PARADEPR}. They take the low dimension CLS vector representations from each passage and learn an additional aggregation model to combine these representations. The aggregation model can any neural model, e.g. a Transformer in PARADE-Transformer~\cite{Li2020PARADEPR}. These pooling systems make it possible to apply deep LMs to document re-ranking. However, they prohibit direct query and full document interaction. The interactions always happen at the passage level and the passage scores or low dimension representations are bottlenecks in the model. 

Alternatively, some research has explored the use of long-text Transformers~\cite{child2019sparsetransformer} that use sparse attention patterns~\cite{Jiang2020LongDR}. Sparse attention comes with the cost of lower capacity than vanilla attention~\cite{child2019sparsetransformer}. 

A third class of approaches tries to first select the important short chunk of document, the ``key block'', and feed it into a vanilla Transformer~\cite{Li2021KeyBLDSK,Kim2021QuerydrivenSS}. These models' success depends on a) the quality of the selected block, and more importantly b) the assumption that a smaller key block exists to represent the full long document.

\section{Methodologies}
In this section, we first give some preliminaries on deep re-rankers based on the Transformer LM. We then briefly discuss the MORES modular re-ranker and how we generalize it for document re-ranking.

\subsection{Preliminaries}
Recent state-of-the-art systems in text re-ranking are based on encoder-only pre-trained Transformer LMs. They cast re-ranking as a sequence pair relation prediction task. Query, document, and a special CLS token are concatenated and fed into the Transformer LM. Query and document go through a series of Transformer layers together, interact, and produce contextualized representations $H$.
\begin{equation}
     H = \text{LM} (\text{concatenate}(\text{CLS};\text{query};\text{document}))
\end{equation}
A final score is computed  as a dot product between the representation of the CLS vector $H_{cls}$ and a projection vector $\mathbf{v}_\text{proj}$.
\begin{equation}
    \text{score} = \mathbf{v}_\text{proj}^\intercal H_{cls}
\end{equation}
Effectively, the Transformer model here serves as a deep non-convex similarity function between query and document. These models, however, have two disadvantages:  1) the aforementioned LM input length constraint, which is typically 512, and, 2) query and document are entangled in a black-box fashion. The former means these models do not natively support long documents. The latter makes it hard to design an easy fix. In this paper, we instead turn to modular re-rankers~\cite{gao-etal-2020-modularized}. 
\subsection{Modular Re-ranker}
The Modular Transformer-based re-ranker, or in short MORES, was introduced by \citet{gao-etal-2020-modularized}. In MORES, query~(prepended with a CLS) and candidate documents are first separately encoded by two Transformer encoders.
\begin{align}
    H^\text{d} &= \text{Encoder}^\text{doc} (\text{document})\\
    H^\text{q} &= \text{Encoder}^\text{qry} (\text{query})
\end{align}
These representations $H^\text{q}$ and $H^\text{d}$ remain independent until they are fed into an Interaction Module~(IM).
\begin{equation}
    H^{in} = \text{IM} (H^\text{q}, H^\text{d}) 
\end{equation}
The IM is a third Transformer that performs self-attention over the query representation $H^\text{q}$. In addition, query-to-document cross attention is performed in each layer. Given an intermediate query representation representation $q$, the cross-attention generates $q'$.
\begin{equation}
    q' = \text{Attend}(q, H^\text{d})
\label{eq:mores-attn}
\end{equation}
This operation interacts query and document and is of complexity linear to query/document length. The final CLS vector is projected to generate a relevance prediction score.
\begin{equation}
    \text{score} = \mathbf{v}_\text{proj}^\intercal H^{in}_{cls}
\end{equation}
\subsection{Modular Long Document Re-ranker}
The flexible design of MORES allows us to further modify the modules to support long documents. Given a long document that does not fit into a typical LM re-ranker, we first break it into chunks, document $\rightarrow \{c_1, .., c_n\}$, and encode each chunk independently.
\begin{equation}
    H^d_j = \text{Encoder}^\text{doc} (c_j) \quad j=1,..,n
\end{equation}
We modify the Interaction Module to take all or the chunks' representations at once,
\begin{equation}
    H^{in} = \text{IM} (H^\text{q}, [H^d_1; H^d_2;..;H^d_n]) 
\end{equation}
where a \emph{single} \textbf{joint query-to-all-chunk cross attention}  operation is performed over concatenated \emph{all} chunks.
\begin{equation}
    q' = \text{Attend}(q, \text{concatenate}(H^d_1; H^d_2;..;H^d_n))
\end{equation}
This joint attention allows flexible query-document token-level interaction: the query can freely attend to segments in the entire document without chunk boundary limitation. This operation is a generalization of the cross-attention operation in \autoref{eq:mores-attn}. Since the full document is only being \emph{attended to} here, this operation remains a complexity linear to document length. We call this new design MORES+, indicating a modification over MORES to support long document re-ranking. 

For familiar readers, we note this design shares a similar spirit with Fusion-in-Decoder~\cite{Izacard2021LeveragingPR} which uses sequence-to-sequence model to combine multiple retrieved contexts for question answering.

\subsection{Complexity and Efficiency Considerations}
\label{sec:comlexity}

\begin{table*}[t]
\centering
\scalebox{1}{
\begin{tabular}{ l | c c c }
\hline
 & Query Dependent & Query Independent & Total  \\
\hline
BERT-maxp & $\mathcal{O}(CD + 2QD + Q^2D/C)$ & - & $\mathcal{O}(CD + 2QD + Q^2D/C)$\\
PARADE-Transformer & $\mathcal{O}(CD + 2QD + Q^2D/C + D^2/C^2)$ & -  & $\mathcal{O}(CD + 2QD + Q^2D/C + D^2/C^2)$ \\
MORES+ & $\mathcal{O}(QD + Q^2)$ & $\mathcal{O}(CD)$ & $\mathcal{O}(CD + QD + Q^2)$\\
\hline 
\end{tabular}
}
\caption{Complexity analysis of attention operations in score-pooling system BERT-maxp, representation aggregation system PARADE-Transformer and our proposed MORES+. This considers a length $D$ document, a length $Q$ query, and length $C$ chunks.}
\label{tab:complexity}
\end{table*}

Here we compare the time complexities of score-pooling systems, representation aggregation systems, and MORES+. We use BERT-maxp~\cite{dai2019deeper} and PARADE-Transformer~\cite{Li2020PARADEPR} as concrete examples. We consider a long document of length $D$, a query of length $Q$, and a chunk size $C$ for all systems. This gives us $\mathcal{O}(D/C)$ chunks. We focus our analysis here on the (self and cross) attention operations' cost of the Transformers. Note that the other major cost factor is the feed-forward network, which is point-wist and remains a cost linear to input length across all systems.

With both score pooling systems~(BERT-maxp) and representation aggregation systems~(PARADE-Transformer), each document chunk is concatenated with the query and separately fed into a Transformer. Self attention costs $\mathcal{O}((C+Q)^2)$ each. A total of $\mathcal{O}(CD + 2QD + Q^2D/C)$ for $\mathcal{O}(D/C)$ chunks.

The aggregator in score aggregation systems has an extra cost. In PARADE-Transformer, self-attention costs $\mathcal{O}(D^2/C^2)$ for latent vectors from $\mathcal{O}(D/C)$ chunks.

In MORES+, each of the documents is encoded independently to the query. Self attention costs $\mathcal{O}(C^2)$ for each chunk, a total of $\mathcal{O}(CD )$. Query encoding costs $\mathcal{O}(Q^2)$. Finally, the interaction module performs cross attention from a length $Q$ query to $\mathcal{O}(D/C)$ of the length $C$ chunks, or equivalently $D$ document tokens. This costs $\mathcal{O}(QD)$ time.

Note that similar to MORES, MORES+ encodes documents independent to the query. These query independent representations can therefore be pre-computed offline. In comparison, both BERT-maxp and PARADE-Transformer start from the concatenated query and document chunks. They, therefore, do not permit any pre-computation. In \autoref{tab:complexity}, we record the complexities of all three systems. Note that a common situation is that a document is a few times longer than a chunk and a chunk is much longer than a query, i.e. $D > C >> Q$. With pre-computation, it is possible to move a costly $\mathcal{O}(CD)$ term offline. On the other hand, when everything is run online, MORES+ has complexity similar to compared systems.

\subsection{Initialization} Encoder-only LMs like BERT~\cite{Devlin2019BERTPO} and RoBERTa~\cite{liu2019roberta} do not perform cross attention and therefore do not contain weights that can initialize the interaction module. In the original MORES paper, the authors propose to borrow self-attention weights but also found the approach to be non-optimal~\cite{gao-etal-2020-modularized}. Observing the emergence of pre-trained sequence-to-sequence encoder-decoder LMs~\cite{Raffel2019ExploringTL,Lewis2020BARTDS}, we propose to instead initialize using their weights. In particular, we initialize with BART~\cite{Lewis2020BARTDS}, using its encoder for our document encoder module and its decoder for our interaction module. The decoder-to-encoder weights are used as MORES+'s query-to-document attention weights. To accommodate this new initialization scheme, we reduce the query encoder module down to a single embedding layer and have queries encoded jointly in the interaction process. In other words, the flows of computation in BART are preserved.\footnote{Details can be found in our open open-source code.}

\section{Experiment Setup}
\subsection{Datasets}
We first verify the effectiveness of MORES+ systems on classical few-shot IR datasets. We use Robust04 and Clueweb09 and report nDCG@20, following \citet{dai2019deeper}. We consider two versions of queries: short keyword queries~(title) and natural language query~(description). An example provided by \citet{dai2019deeper} is shown in \autoref{tab:robust-example}. 

We also adopt the MS-MARCO document ranking dataset~\cite{Campos2016MSMA}. MS-MARCO document ranking is a recent popular large-scale supervised dataset constructed from Bing with substantially more training queries. Its leaderboard has received submissions from a variety of production-ready systems. Participants are given Train/Dev sets judgments and the submissions are evaluated based on the hidden Eval set performance. MS-MARCO helps demonstrate MORES+'s ability to handle contemporary real-world web search tasks. Queries in MS-MARCO document ranking are of natural language form. We report the official metric MRR@100 on Dev and Eval~(leaderboard) query sets.  
\begin{table}[t]
\centering
\scalebox{0.9}{
\begin{tabular}{ c l}
\hline
\hline
Title &  air traffic controller \\
Description & \multirow{2}{7cm}{What are working conditions and pay for U.S. air traffic controllers?} \\
& \\ 
\hline \hline
\end{tabular}
}
\caption{Example of Robust04 search topic (Topic 697).}
\label{tab:robust-example}
\end{table}
\begin{table}[t]
\centering
\scalebox{1}{
\begin{tabular}{ l | c c c }
\hline
 & \# Queries & \# Docs & \# Tokens / Document  \\
\hline
Robust04 & 249 & 0.5 M & 0.7K\\
ClueWeb09 & 200 & 50M & 3.3K \\
MS-MARCO & 0.37M & 3.2M &1.3K\\
\hline 
\end{tabular}
}
\caption{Collection statistics.}
\label{tab:data-stats}
\end{table}

We show the three datasets' statistics in \autoref{tab:data-stats}. Note that all three have documents on average longer than the typical LM input length of 512. They help us test the effectiveness of long document re-rankers.

\subsection{Implementations}
Our MORES+ models are implemented based on BART code in the Huggingface transformer library~\cite{Wolf2019HuggingFacesTS}. The document encoder and interaction module both have 12 layers. We set each input chunk size to 512, similar to the maximum input sequence length of BART during pre-training~\cite{Lewis2020BARTDS}. We use a maximum of 3 chunks on Robust04 and MS-MARCO and a maximum of 6 chunks on ClueWeb09, preserving up to $\sim1.5$k tokens and $\sim3$k tokens in each input respectively. The models are trained with localized contrastive estimation (LCE) loss~\cite{Gao2021RethinkTO}. Following previous work~\cite{Li2020PARADEPR,Jiang2020LongDR}, we first pre-train MORES+ on MS-MARCO passage ranking data to warm up the model. \footnote{Training details are available in our open-source code.}   

We use the commonly adopted Indri initial rankings released by \citet{dai2019deeper} on Robust04 and Clueweb09. Since one of our major baselines, the previous state-of-the-art system PARADE uses a different set of BM25+RM3 initial rankings on Robust04, we also use MORES+ to re-rank BM25+RM3 initial rankings on Robust04 released by \citet{Yang2019CriticallyET}. We perform 5-fold cross-validation on Robust04 and ClueWeb09 as described in \citet{dai2019deeper}.

On MS-MARCO, we follow top leaderboard systems and use a dense initial stage retriever. Our dense retriever is fine-tuned from a pre-trained coCondenser~\cite{co-condenser}. It combines a special dense pre-training architecture Condenser~\cite{Gao2021CondenserAP} and contrastive learning to pre-train a tailored model for dense retrieval. It allows us to perform a simple low-cost fine-tuning. For the leaderboard submission, we ensemble 4 models that are trained with different initialization.

\subsection{Baseline Systems}
On the classical few-shot IR datasets, we compare with score pooling systems BERT-firstP/maxP/sumP~\cite{dai2019deeper}, and latent representation aggregation system PARADE-CNN/Transformer~\cite{Li2020PARADEPR}. We also consider long-text Transformers. QDS-Transformer~\cite{Jiang2020LongDR} uses a special attention pattern for ranking. A few universal long-text Transformer baselines, Sparse-Transformer, Longformer, and Transformer-XH are also taken from \cite{Jiang2020LongDR}. A recent method BeST~\cite{Kim2021QuerydrivenSS} that uses query-based key-block selection is also considered. We also borrow some earlier classical baseline systems from \citet{dai2019deeper} including SDM, RankSVM and Coor-Ascent as well as pre-BERT neural models DRMM and Conv-KNRM. Due to the high training cost of these long-text models, we directly copy previously reported numbers.

On MS-MARCO, we report the performance of the current top leaderboard systems, as shown in \autoref{tab:marco-doc}. Many of them are or are based on real-world production-ready systems~\cite{Xiong2021ApproximateNN,Zhang2022UniRetrieverTL}.


\section{Experiment Results}

\begin{table}[t]
\centering
\scalebox{0.9}{
\begin{tabular}{ l || c  c | c c }
\hline \hline
& \multicolumn{4}{c}{nDCG@20} \\
& \multicolumn{2}{c|}{\textbf{Robust04}} & \multicolumn{2}{c}{\textbf{ClueWeb09}} \\
Model & Title & Description & Title & Description \\
\hline
\textbf{Classical Systems} & & & & \\
SDM & 0.427 & 0.427 & 0.279 & 0.235 \\
RankSVM & 0.420 & 0.435 & 0.289 & 0.245\\
Coor-Ascent & 0.427 & 0.441 & 0.295 & 0.251\\
\hline
\textbf{Neural Systems} & & & & \\
DRMM & 0.422 & 0.412 & 0.275 & 0.245 \\
Conv-KNRM & 0.416 & 0.406 & 0.270 & 0.242 \\
Sparse-Transformer & 0.449 & - & 0.274 & - \\
Longformer-QA & 0.448 & - & 0.276 & - \\
Transformer-XH & 0.450 & - & 0.283 & - \\
BERT-firstP~\cite{dai2019deeper} & 0.444 & 0.491 & 0.286 & 0.272    \\
BERT-maxP~\cite{dai2019deeper} & 0.469 & 0.529 & 0.293 & 0.262    \\
BERT-sumP~\cite{dai2019deeper} & 0.467 & 0.524 & 0.289 & 0.261 \\
QDS-Transformer~\cite{Jiang2020LongDR} & 0.457 & - & 0.308 & - \\
BeST~\cite{Kim2021QuerydrivenSS} & 0.487 & 0.537 & - & - \\
PARADE~\cite{Li2020PARADEPR} & &  &  & \\
 \;- CNN$^\text{RM3}$ & 0.563 & 0.610 & - & - \\
 \;- Transformer$^\text{RM3}$ & 0.566 & 0.613 & - & - \\
\hline
MORES+ & 0.556 & 0.612 & \textbf{0.353} & \textbf{0.329} \\
MORES+$^\text{RM3}$ & \textbf{0.569} & \textbf{0.641} & - & - \\
\hline \hline 
\end{tabular}
}
\caption{Results on Robust04 and ClueWeb09.  Results not available are denoted with `-'. RM3: These systems use BM25+RM3 initial rankings~\cite{Yang2019CriticallyET} instead of Indri initial rankings~\cite{dai2019deeper}.}
\label{tab:rb-cw}
\end{table}
\subsection{Classical Few-shot IR}
\autoref{tab:rb-cw} shows the performance of the baseline systems and MORES+ on Robust04 and ClueWeb09. We see that MORES+ achieves the new state-of-the-art in ranking effectiveness. 

On Title queries, many neural re-rankers, despite being of higher capacity, do not show a clear advantage over lighter classical systems. Recall that these Title queries are typically of one or a few words, which are simple but ambiguous. As argued by \citet{dai2019deeper}, they typically require less of the language understanding that neural models excel at. Similarly, these keyword-like queries do not require sophisticated query document interaction. In our experiments, we see on Robust04, the best PARADE systems, which pool passage latent vectors into global relevance scores, have similar performance to MORES+\footnote{PARADE results should be compared with MORES+$^\text{RM3}$ entries, the BM25+RM3 results which use the same initial retriever.}. 

On the other hand, when moving over to Description queries that are written in natural language, neural models start to show their advantages on language understanding. In particular, models that are based on pre-trained LMs have a clear performance margin over classical systems. Meanwhile, the direct interaction between the query and the full long document also proves to be of decent utility. We see that on both Robust04 and ClueWeb09 Description queries, MORES+ has a clear advantage over all previous methods, including score-pooling systems and latent representation pooling systems. 
As many search engines/portals are taking up more user-friendly natural language form queries, we believe the improvements brought by MORES+ will have further future impacts. 

We also see that MORES+ can outperform sparse Transformer based models. Both sparse Transformer re-rankers and MORES+ attempt to preserve granular query-document interaction. Sparse Transformers re-rankers do so by making the self-attention pattern sparse, effectively lowering its cost at the cost of capacity. As discussed by \citet{Li2020PARADEPR}, these unique patterns of sparse attention are not optimal for re-ranking. In comparison, MORES+ uses another form of query-to-document attention which shows better effectiveness here. On the other hand, sparse Transformers are more efficient than other models. Note that, as discussed in \autoref{sec:comlexity}, modular re-rankers support document representation pre-computations, which can substantially speed up re-ranking by tens of times. We focus on effectiveness impacts in this paper and leave the exploration of efficiency to future work.

In addition, MORES+ also outperforms the key block method BeST. We can conceptually think of the cross attention operation in MORES+ as a soft information selection process. It can freely attend to useful pieces in the document, by assigning higher attention weights. This soft process does not put hard constraints nor require the pieces to be contiguous. In other words, MORES+ can more flexibly select ``key information'' beyond ``key block''.

\begin{table}[t]
\centering
\scalebox{0.85}{
\begin{tabular}{ l || c c }
\hline \hline
\multicolumn{3}{c}{\textbf{MS-MARCO Document Ranking Leaderboard}}\\
\hline
& \multicolumn{2}{c}{MRR@100} \\
Model & Dev & Eval \\
\hline
UniRetriever & 0.500 & \textbf{0.440} \\
Group-HNS-Retrieval+Multi-Granularity-Rerank & 0.496 & 0.436 \\
ANCE MaxP + LongP / SEED-Encoder+LongP (ensemble) & 0.487 & 0.427 \\
hybrid retriever / misc. BERT-longp & 0.481	& 0.424 \\
PROP\_step400K base + doc2query top1000(ensemble v0.2) & 0.479 & 0.423 \\
\hline
coCondenser + MORES+ & \textbf{0.501} & 0.436\\
- single (no ensemble) & 0.493 & - \\
\hline \hline 
\end{tabular}
}
\caption{Results on MS-MARCO document ranking. Top 5 compared runs as of Jan 15th 2022 are recorded.}
\label{tab:marco-doc}
\end{table}
\subsection{Large-scale Supervised IR}
In \autoref{tab:marco-doc}, we show performance on the MS-MARCO document ranking dataset. We see that, at the time when this paper is written, MORES+ achieves 2nd best performance on the Eval set and the best on the Dev set. Early work~\cite{Gao2021RethinkTO} on the MS-MARCO document ranking has shown that a) it requires stronger initial retrievers than BM25, but b) stronger retriever rankings are harder to re-rank. This means, for our discussion, it is important to verify the re-ranker's improvement when paired with a strong retriever. Here we see that MORES+ can successfully re-rank candidates produced by one of the state-of-the-art dense retrievers, achieving competitive results. 

When making further comparisons with other runs, we would like to remind our readers that other leaderboard systems are typically heavily engineered systems that combine various techniques. In comparison, ours is a simple two-stage system with a pair of dense retrievers and a modular re-ranker. In the last row of \autoref{tab:marco-doc}, we show Dev set performance of the MORES+ system without ensemble. It can still achieve competitive results and outperform several top leaderboard systems. 

\subsection{Ablation: Number of Input Chunks}

In this section, we consider an ablation study where we input smaller numbers of chunks to MORES+. Ideally, as the number of chunks increases, a good model should be able to pick up more information and generate better final relevance judgments. It should not be confused by extra noise in the input if the added chunks are not relevant. On the other hand, even with less sufficient information, the model should not fail catastrophically but retain a certain level of performance.
In practice, reducing the number of input chunks can help lower the total computation amount. It helps lower search latency when search volume increases.  

\begin{figure}[t]
  \centering
  \begin{subfigure}[b]{0.47\textwidth}
  \includegraphics[width=0.99\textwidth]{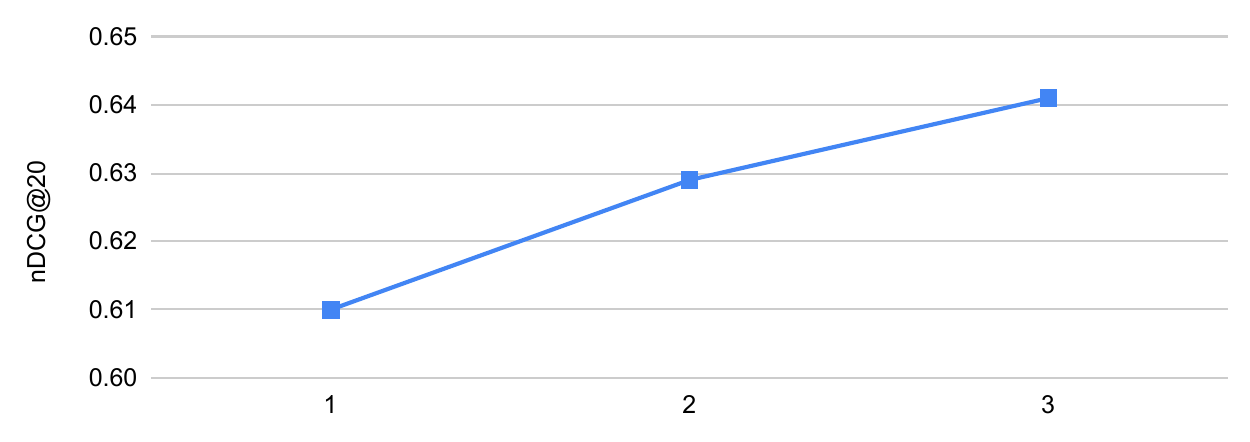}
  \caption{Robust04 performance measured with nDCG@20.}
  \end{subfigure}
  \begin{subfigure}[b]{0.47\textwidth}
  \includegraphics[width=0.99\textwidth]{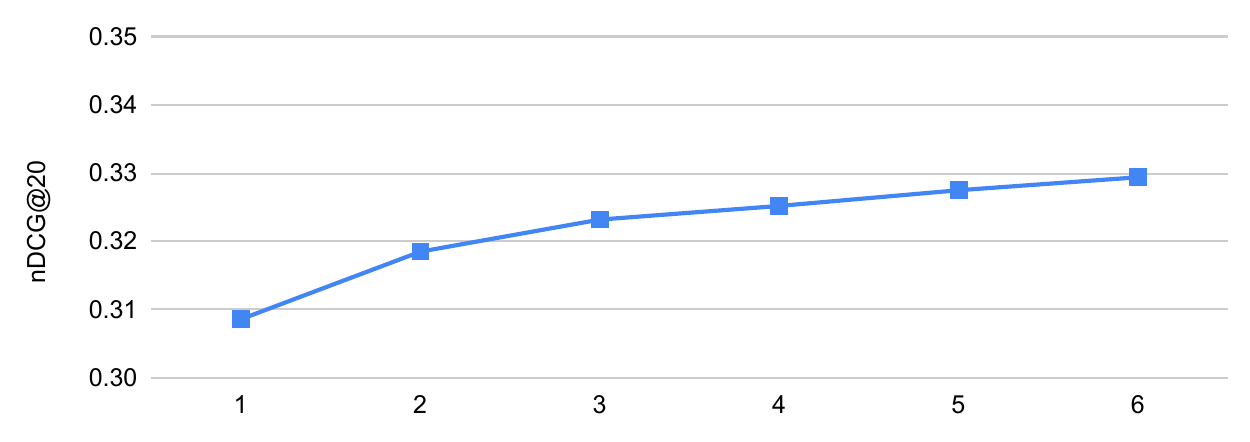}
  \caption{ClueWeb09 performance measured with nDCG@20.}
  \end{subfigure}
  \caption{Effect of changing number of input chunks to MORES+. Robust04 has an average length of 0.7K and ClueWeb09 an average length of 3.3K. We used up to 3 chunks on Robust04 and up to 6 on ClueWeb09.}
  \vspace{-0.1mm}
  \label{fig:ablation}
\end{figure}

In \autoref{fig:ablation}, we plot number of input chunks versus nDCG@20 on Robust04 and ClueWeb09 using description queries. Note here a special case is the single chunk situation, where MORES+ reduces to the original MORES system. This corresponds to the left-most data point in each plot. Here, MORES+ shows monotonic improvements with respect to the number of input chunks while remaining decently effective with a lower number of chunks.

\section{Conclusion}
In this paper, we present MORES+, a new approach for re-ranking long documents. It deviates from previous long document systems that rely on an encoder-only Transformer and is based on a modular re-ranker framework, MORES. We introduce a granular token-level query document interaction method using query to document chunk cross attention. We use this method to augment the original MORES framework for long document re-ranking. 

Our experiments demonstrate the effectiveness of MORES+ on both classical IR datasets with few training examples and contemporary large-scale supervised datasets. We found MORES+'s new form of token-level interaction is especially beneficial for natural language queries. It achieves a new state-of-the-art on Robost04 and ClueWeb09 using Description queries, with large improvements over baselines. On MS-MARCO document ranking, MORES+ also shows competitive performance, with second-best Eval and best Dev set performance.

We believe MORES+ opens new possibilities in document re-ranking using modular re-rankers. Broadly, it reminds people that instead of sticking with the original Transformer encoder architecture used in models like BERT, document ranking can also be solved with innovations in architecture. On the other hand, MORES+ shows competitive performance to real-world systems. Future work can explore the possibility of using MORES+ in real search tasks.

\section*{Acknowledgments}
This work was supported by the National Science Foundation (NSF) grant IIS-1815528.  Any opinions, findings, and conclusions in this paper are the authors’ and do not necessarily reflect those of the sponsor. The authors would like to thank Google’s TPU Research Cloud~(TRC) for
access to Cloud TPUs and the anonymous reviewers for the reviews.

\clearpage
\bibliographystyle{ACM-Reference-Format}
\balance
\bibliography{ref}

\end{document}